# AN INTEGRATED FRAMEWORK FOR UNCERTAINTY QUANTIFICATION IN HIGH TEMPERATURE GAS COOLED REACTORS USING THE HCP TIME-DEPENDENT MULTIPHYSICS CODE AND DAKOTA TOOLKIT

W. Osman[1], A. Sadek[1], M. R. Altahhan[1], C. Liu[1,2], M. Avramova[1], K. Ivanov[1]

[1] Department of Nuclear Engineering, North Carolina State University, Raleigh, NC 27695, USA
[2] Chair of Nuclear Technology, Technical University of Munich, 85748, Garching, Germany

wiosman@ncsu.edu, atsadek@ncsu.edu, mraltahh@ncsu.edu, cliu49@ncsu.edu, mnavramo@ncsu.edu, knivanov@ncsu.edu

## ABSTRACT

The High Temperature Reactor Code Package (HCP) provides sophisticated modeling and simulation capabilities for high temperature gas cooled reactors (HTGRs) like the HTR-200 Modul. However, HCP currently lacks integrated methods for uncertainty quantification (UQ) and sensitivity analysis. This work aims to couple HCP with the DAKOTA toolkit to enable UQ workflows for quantifying how different uncertainties impact HTGR system performance. DAKOTA offers state-of-the-art sampling and analysis methods that will be linked to the HCP time-dependent multiphysics environment. Key input parameters related to manufacturing variability, boundary and initial conditions, and material properties will be defined as uncertain in this study. Both steady state and time-dependent multiphysics simulations will be analyzed to understand the relative importance of uncertainties across different physics phenomena. Output metrics of interest can include anything measured by the HCP code (e.g., criticality, maximum fuel temperature, power profiles, among others). Results show the HTR-200 Modul design's robustness to input uncertainties related to inlet gas temperature, U-235 enrichment, graphite density, inlet mass flow rate, and reactor power. A pressurized loss of forced cooling transient was simulated against uncertainty in the inlet gas temperature. Results show the maximum fuel temperature is within design limits with a relatively big safety margin even when considering uncertainty in the boundary conditions of the reactor. Future work will expand the analysis to more HCP physics modules and nuclear data uncertainties. By leveraging UQ and sensitivity analysis with the linked HCP/DAKOTA environment, this work will provide valuable new insights into the expected performance variability of HTR systems during normal and off-normal conditions. The framework developed here will aid uncertainty management activities by gas-cooled reactor developers, researchers, and regulators.

## 1. INTRODUCTION

High temperature gas-cooled reactors offer enhanced safety and efficiency versus traditional reactor designs. However, quantifying uncertainty in the complex multiphysics performance of HTGRs remains a challenge for licensing and development. The HCP code [1] provides sophisticated modelling capabilities that can allow one to model and simulate HTGR type reactors. When matched with UQ methods, one can reliably bound outcomes (i.e., responses and results) based on the uncertainty on the inputs. This work will leverage the Design Analysis Kit for Optimization and Terascale Applications (DAKOTA) UQ software [2] to link UQ workflows with HCP for *probabilistic* simulation of the HTR-200 Modul HTGR reactor which is the basis for the current HTR-PM design under operation in China [3]. The multiphysics interactions between neutronics, heat transfer, and fluid flow lead to inherent uncertainties that must be characterized. By propagating uncertainties in key inputs through HCP and the developed UQ framework, this research aims to provide reactor developers, scientists, and regulators with critical insights into where resources should be allocated to improve confidence in HTGR predictions and constrain performance uncertainties. The UQ framework implemented here can also serve as a framework for analysis of other HTGRs toward licensing and design activities.

As described in Section 2, the developed framework will be introduced alongside the HCP code. In Section 3, we demonstrate application of this framework on key uncertainties in multiphysics and time dependent exercises using the HTR-Modul-200 reactor. Finally, Section 4 summarizes conclusions and discusses future work to expand this coupled UQ-framework to cover nuclear data uncertainty propagation, and other modules available inside HCP (e.g., dust source term analysis).

## 2. FRAMEWORK FOR HTGR UQ PROPAGATION USING HCP AND DAKOTA

### 2.1 DAKOTA Part of the Framework

Central to the UQ framework is a DAKOTA input file that controls and executes the analysis. This input file calls a Python driver script to manage the workflow interactions between DAKOTA, HCP, and the supporting pre/post-processing routines. Users first develop template versions of the HCP input files needed for the simulations, using descriptor placeholders for parameters that will be uncertain. These descriptors identify the variables that will have values sampled from defined probability distributions. Prior to each HCP simulation, the Python script and DAKOTA-generated parameters file update the input template, swapping descriptors for the sampled variable values. This input preparation utilizes pre-processing functions to enable propagating the sampled variables into the physics simulations executed by HCP. The templated workflow allows efficient linking of the sampling and analysis strengths of DAKOTA with the multiphysics capabilities of HCP for uncertainty propagation.

DAKOTA provides an extensive selection of sampling and analysis techniques, making it a flexible tool for uncertainty quantification. It can account for both epistemic uncertainties due to incomplete knowledge and aleatory uncertainties inherent from random effects. To assess the impacts of these two types together, DAKOTA can utilize a nested sampling approach [4]. This involves distinct outer and inner sampling loops to separately propagate the epistemic and aleatory uncertainties while capturing their interactions. In the outer loop, a sampling method (e.g., Latin Hypercube sampling) draws values for the epistemic parameters across their possibility distributions. Then, for each outer sample, the inner loop samples the aleatory input variables based on their inherent probability distributions. This propagates stochastic variability through each simulation instance. By sampling aleatory uncertainties, the method generates confidence intervals and distributions reflecting output performance variability. A key benefit of nested sampling is producing probability bounds that incorporate epistemic uncertainty impacts on reliability despite aleatory variations. Dakota aggregates statistics like percentile confidence intervals across outer samples into a probability box, visualizing the combined effect of both uncertainty types on outcomes when needed.

DAKOTA executes many HCP simulations, sampling variable values based on specified probability distributions configured in the DAKOTA input file. For each case, DAKOTA prepares working directories containing the sampled variable sets. The Python driver script substitutes the sampled values in place of descriptors in the HCP input templates. After the multiphysics simulation is completed in HCP, a post-processing script extracts the defined response metrics (criticality, temperature, power, etc.) to output files. DAKOTA then performs statistical analyses on the responses to characterize uncertainty propagation through the physics models. The user can also perform independent analysis from DAKOTA by inspecting and extracting relevant information from the HCP outputs. This integrated framework enables assessing the impacts of uncertainties on simulated performance. The modular workflow allows customization to new problems by altering the DAKOTA sampling specs, HCP input templates, and post-processing scripts. This flexible linkage of sampling, multiphysics simulation, and analysis provides a platform to gain key insights into reliability and robustness of HTGRs. This framework is depicted in Figure 1.

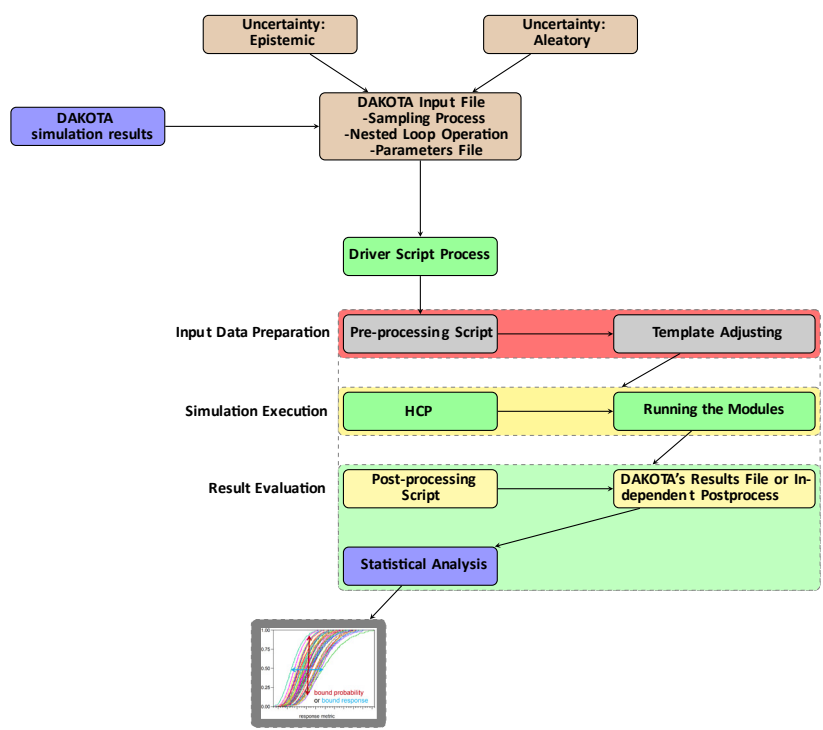

**Figure 1: The UQ Framework between DAKOTA and HCP.**

## 2.2   HTGR Modelling and Simulation with the HCP Code

HCP is a sophisticated modeling and simulation software platform tailored to analyzing HTGRs. HCP was developed through a joint research project between the Institute for Reactor Safety and Reactor Technology at RWTH Aachen University and the Institute for Energy and Climate Research at Forschungszentrum Jülich. The code system integrates modeling capabilities across relevant physics including neutronics, heat transfer, and fluid flow to enable multiphysics calculations of HTGR performance. HCP consists of an array of specialized modules linked through an integrated C++ backbone structure as seen in Figure 2. Key modules focus on fluid dynamics (MGT-FD), 3D neutron diffusion (MGT-N), burnup (TNT), fission product transport (STACY) including dust particles resuspension (STAR), spectrum analysis (TRISHA), and fuel management (SHUFLE). Both MGT-FD and MGT-N are based on the advanced HTGR simulation code MGT-3D, which is the successor and the 3D multigroup version of TINTE [1, 5]. The HCP backbone, written in C++, coordinates data flows between modules (that are written in modern Fortran) and contains a nuclear data library generator for providing consistent cross section data for the neutronics simulations. HCP's flexibility supports modeling steady state conditions, transients, and accident scenarios in both pebble bed and prismatic block type HTGR designs. Advanced programming techniques facilitate efficient data exchange for integrated multiphysics analyses not practical with standalone codes. HCP output provides details on critical reactor performance metrics under a range of operating conditions and accident initiators.

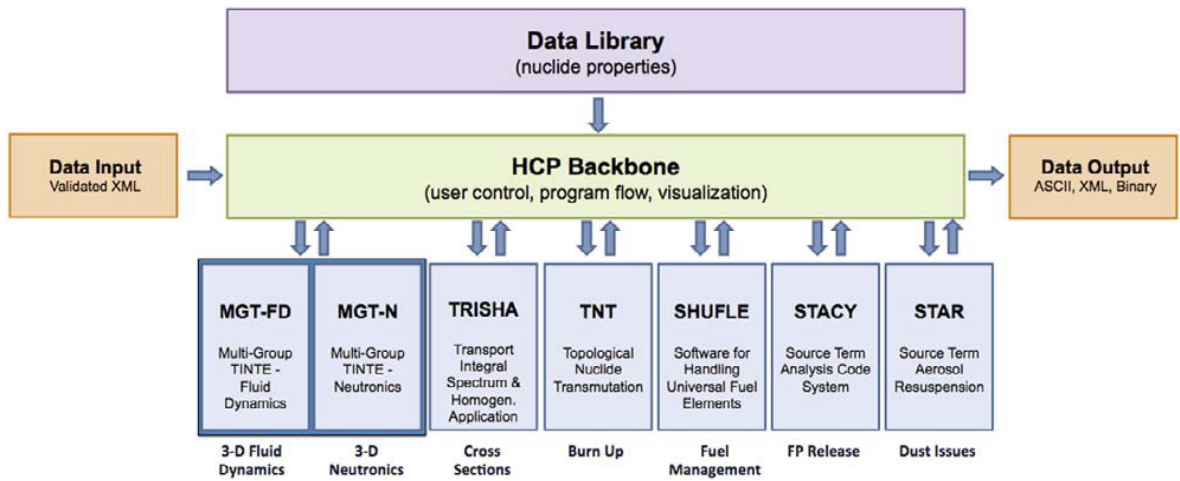

**Figure 2: Chart Depicting the Different Modules and Codes found inside the HCP M&S Package [1].**

At its core, HCP coordinates coupled calculations across three main physical phenomena - neutronics, heat transfer, and fluid dynamics (which includes a corrosion analysis and chemical reactions simulation module). The neutronics module, MGT-N, leverages a 3D neutron diffusion solver (based on the 2D/1D leakage iterative method [1]) to provide the spatial flux and fission power distribution. This couples to the thermal-hydraulics module MGT-FD based on a porous media model for the gas coolant and graphite moderator [5]. The integrated data backbone enables seamless transfer of temperature and power data between modules to capture underlying physics interdependencies. The input file system in HCP is modular and generally contains 3 input files in XML language. Additional input files can be used depending on the analysis type (e.g., generation of fine structure nuclear data library using NJOY [6]). The 3 input files contain one for describing the location of data files, nuclear data libraries, and some general information on the simulation. The other two input files are for the model description and the scenario being simulated (e.g., details of the steady state or transient simulations). Since both the model and the scenario files can be analyzed for uncertainty propagation, the python script driver for the developed UQ framework is flexible to accept both input files as sources of input parameters to be perturbed (and in future work, additional inputs for nuclear data library generation).

## 3. DEMONSTRATION OF THE DEVELOPED UQ FRAMEWORK

Three scenarios will be examined in this paper, and they cover both steady state and time-dependent scenarios for single and multiphysics simulations. These simulations will include a burnup analysis using MGT-N and TNT modules of the HCP code package, and steady state and time-dependent multiphysics simulations using the coupled MGT-N and MGT-FD codes. All the simulations will use the 2D model of HTR-Modul-200. A key aim for this wide range of scenarios is to demonstrate the successful integration of the framework with HCP, showcasing its capabilities for *reactor's dynamics* simulations. The 2D R-Z section of the HTR-Modul-200 is shown in Figure 3. The reactor has a nominal thermal power of 200 MWth, producing 80 MWe of electrical power. The pebble bed geometry is cylindrical with an average height of 9.43 m and a diameter of 3 m. Operating conditions include a coolant inlet temperature of 250°C, and an outlet temperature of 700°C, with an average core pressure of 60 bar, and top-to-bottom coolant flow through the pebble bed. The $UO_2$-based fuel pebbles are enriched to 8% and contains around 12,000 TRISO-coated particles. The 2D model of the HTR-MODUL-200 encompasses crucial components such as the pebble bed, cavity, upper, bottom, and side reflectors, upper, bottom, hot gas plenums, as well as pipes and ducts. Each of these have its own set of equations and correlations inside HCP which uses a porous media approach to model the fluid dynamics. The reactor is described in detail in reference [5].

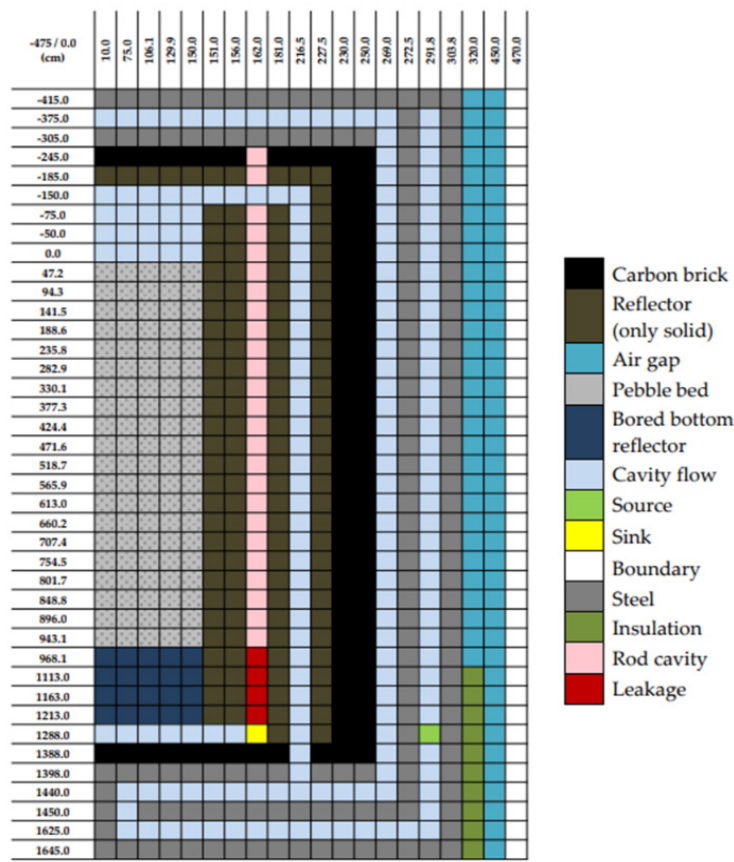

**Figure 3: The HTR-Modul-200 Model Depicting Different Media and Materials inside the Model [5].**

### 3.1    Scenarios Being Studied

For the multiphysics simulation, three scenarios are explored to understand the implications of various operational conditions on the HTR-Modul-200 reactor's performance, focusing on key Figures of Merit (FOMs) such as multiplication factor ($k_{eff}$), fuel temperature, and power output. The first scenario presents a steady-state condition, initiated with initial conditions derived from an equilibrium state of the HTR-Modul-200 reactor. This scenario simulates predetermined reactor operation conditions, providing a baseline for reactor performance metrics. The second scenario transitions into the domain of transient analysis, commencing at an equilibrium state with a nominal power of 200 MW and a gas inlet temperature of 250°C. Initially, a quasi-stationary transient phase lasts for 2 minutes without any changes to the input parameters, representing a stable operational condition. This phase is followed by a transient event, in which the inlet temperature is ramped up from 250°C to 300°C at a rate of 25°C/min as shown in Figure 4. This simulates the effect of a reduced opening cross-section of the feed-water system's valve (found in the secondary loop in the original HTR-Modul-200 design) [5]. The scenario concludes with the reactor power stabilizing, after which a steady-state calculation is performed using the transient's final conditions to evaluate the reactor's return to equilibrium and its ability to handle such thermal stress. Another scenario under the time-dependent multiphysics analysis is the sudden loss of flow cooling under pressure. In this scenario, the mass flow rate of the cooling fluid (the helium gas) is reduced to zero over 30 seconds. This scenario is set to be similar to the pressurized loss of forced cooling accident (PLOFC). Both transients' scenarios are depicted in Figure 4. The primary focus for both transient scenarios is on monitoring the evolution of reactor power and maximum fuel temperature during the simulations.

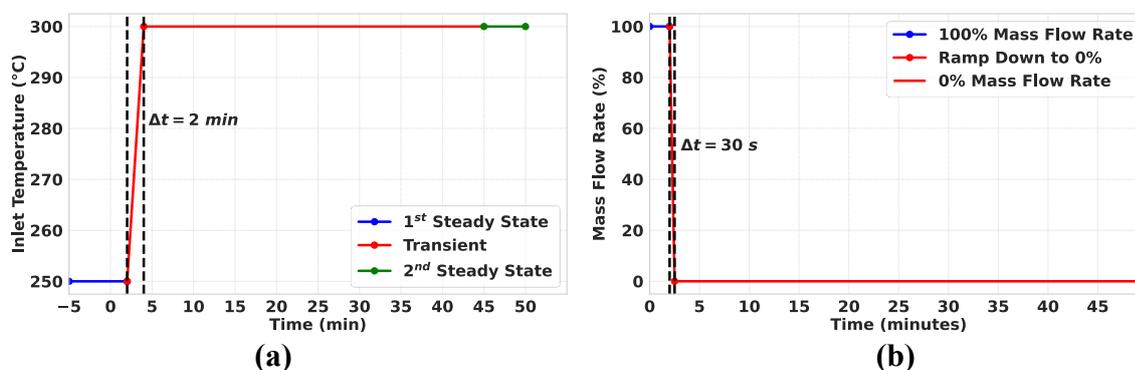

**Figure 4: The Different Transient Scenarios modelled for the HTR-Modul-200 (a) Feed-Water valve partial blockage (b) Pressurized loss of forced cooling.**

Lastly, a neutronics case designed for a depletion analysis is considered. In this scenario, no fluid dynamics is involved, and the core contains only fresh fuel. Only the active core and the surrounding reflector portions of the HTR-Modul 200 are modelled. The transient progresses through 5-day burnup steps over a duration of 30 effective full power days (30 EFPD), incorporating steady-state calculations at the beginning of each time step to capture the reactor's neutronic behaviour over the depletion period.

### 3.2 Sources of Uncertainties and Distribution Types

For the uncertainty analysis within both the multiphysics and neutronics cases, different uncertain input parameters can be selected. These parameters are critical for understanding the impact of variability on reactor performance and safety. The multiphysics cases focus on variables such as inlet gas temperature, inlet mass flow rate, and reactor power, each associated with uncertainty that reflects operational variability, and measurement inaccuracies. In the neutronics case, the analysis extends to parameters such as initial U-235 fuel enrichment and graphite density, which influence the reactor's neutronic behaviour and fuel cycle efficiency. Table 1 and Table 2 detail the means, standard deviations, and distribution types for these parameters. The chosen distributions aim to represent the expected variability in these parameters and are chosen from several references in the literature [7-9].

**Table 1: Multiphysics Uncertainty Study Input Parameters.**

| Parameter | Mean value | 2σ value | Relative 2σ value | PDF type |
|---|---|---|---|---|
| Reactor inlet gas temperature | 250ºC | ±10ºC | ±4% | TN[a] |
| Inlet mass flow rate | 85.45 kg/s | ±3.418 kg/s | ±4% | TN |
| Reactor power | 200 MW | ±20 MW | ±10% | TN |

(a) TN: Truncated Normal Distribution

**Table 2: Neutronics Uncertainty Study Input Parameters.**

| Parameter | Mean value | 2σ value | Relative 2σ value | PDF type |
|---|---|---|---|---|
| Initial fuel enrichment | 8% | ±0.08% | ±1% | TN |
| Graphite density (Reflector and matrix) | 1.75 g/cm$^3$ | ±0.00105 g/cm$^3$ | ±0.06% | TN |

The uncertainties in the multiphysics and neutronics cases are assumed to follow a Truncated Normal (TN) distribution for all parameters (truncated at 2σ values from the mean). This approach ensures that variations remain within realistic bounds, facilitating accurate assessments of reactor behavior and safety across operational scenarios. The input parameters chosen in Tables 1 and 2, which cover initial/boundary conditions, manufacturing uncertainties, and material properties, represent only a subset of the parameters that could potentially be analyzed using the developed UQ framework. The framework permits users to incorporate any uncertainties related to manufacturing, boundary conditions, or material properties. The specific sources of uncertainty used herein were selected because uncertainty ranges for these parameters are available in the HTGR literature, enabling their use for uncertainty propagation [7-9]. By focusing on parameters with established uncertainty ranges, this work demonstrates the capability to propagate input uncertainties through the model; however, the framework's flexibility (allowed by both HCP and DAKOTA) allows extension to other sources in future work like those related to nuclear data or other sources.

### 3.3 Results and Analysis

#### 3.3.1 Steady state multiphysics case

Each of the uncertainty sources in Table 1 was used independently in the uncertainty study. The empirical cumulative distribution function (ECDF) is used to assess the effect of each quantity on both the multiplication factor and the maximum fuel temperature. The HCP source code was modified to enable the ready retrieval of FOMs from the output file for analysis by using dedicated post processing scripts and tools. A sample size of 200 was used for each uncertainty source. Figures 5-7 display the cumulative distribution functions (CDFs) and empirical CDFs for $k_{eff}$ and maximum fuel temperature, respectively, along with Dvoretzky-Kiefer-Wolfowitz (DKW) confidence bounds. The normal distribution CDFs are plotted based on assuming a normal distribution with the mean and standard deviation computed from the 200 samples. The results show the ECDFs follow the normally distributed CDFs, suggesting the output data follow a normal probability density function, as verified by a histogram plot. The narrow distributions signify stable operation even when perturbing the different uncertainty sources.

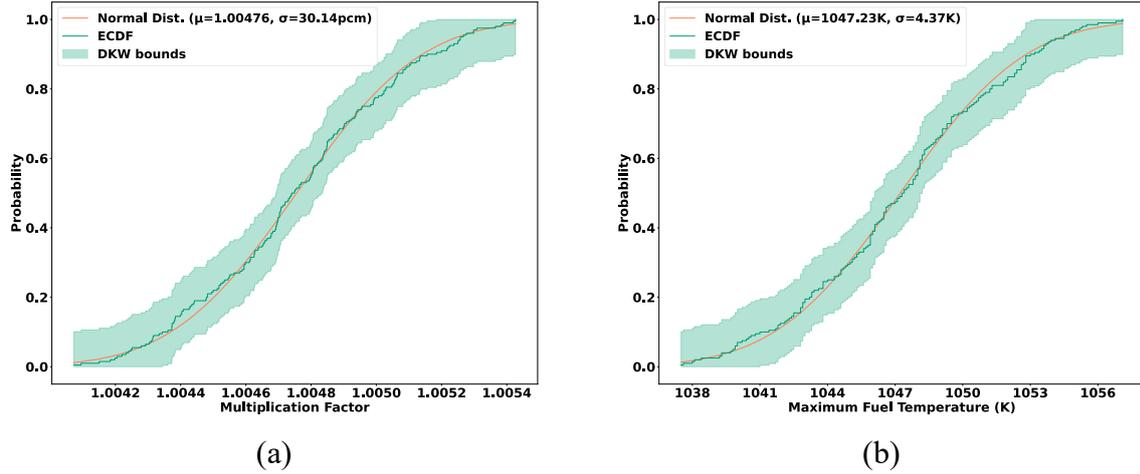

(a)            (b)

**Figure 5: CDFs and ECDFs for (a) Multiplication Factor and (b) Maximum Fuel Temperature under Perturbation of Inlet Gas Temperature.**

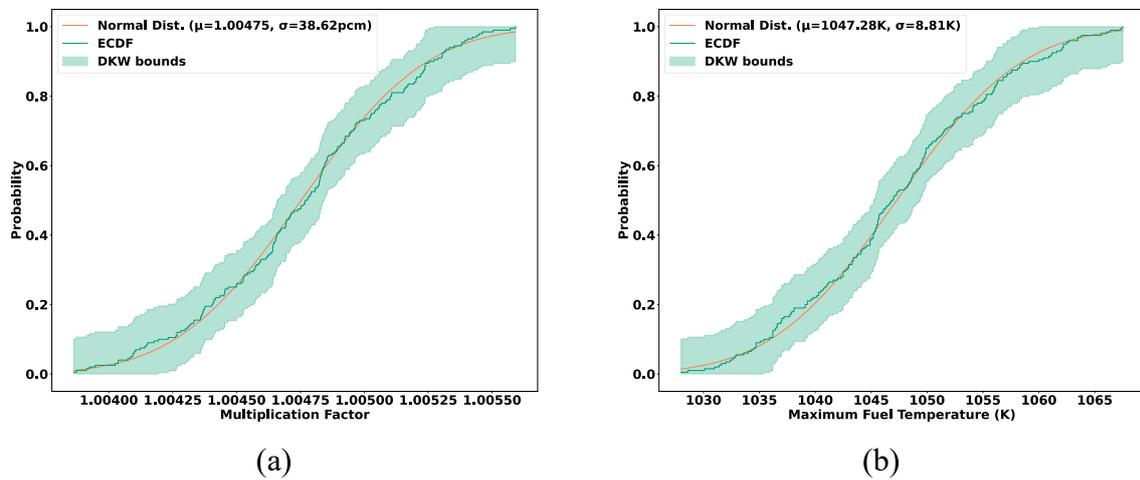

(a)            (b)

**Figure 6: CDFs and ECDFs for (a) multiplication factor and (b) Maximum Fuel Temperature under Perturbation of Inlet Mass Flow Rate.**

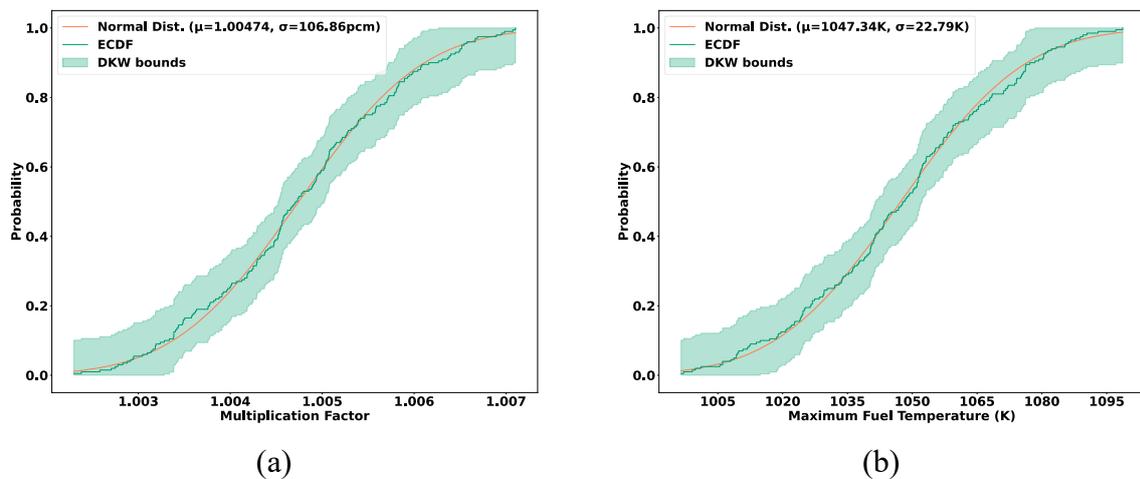

(a)            (b)

**Figure 7: CDFs and ECDFs for (a) Multiplication Factor and (b) Maximum Fuel Temperature under Perturbation of Reactor Power.**

Despite the different sources of uncertainty, the mean values of the output data remain roughly consistent among different sources of uncertainty. This suggests that the central tendency of the system's response is robust to these uncertainties. However, the standard deviations of the output data vary, indicating differing levels of variability in the system's response to each uncertainty source. E.g., the effect of power perturbation exhibits a larger standard deviation compared to the effect of inlet mass flow rate or the inlet gas temperature perturbations, implying a wider range of outcomes and hence a more pronounced effect. This observation shows the importance of not only considering the mean response but also the variability when assessing the impact of different uncertainty sources. It also highlights the value of using tools like ECDFs and DKW bounds in uncertainty analysis, as they provide a more comprehensive view of the system's behavior under uncertainty. These insights can guide the prioritization of efforts in uncertainty reduction, ultimately contributing to the enhancement of system reliability and performance.

### 3.3.2 *Transient multiphysics case*

For the first transient scenario, the focus shifts to the reactor's response to a surge in inlet gas temperature. This transient scenario, in which the inlet gas temperature is ramped from 250°C to 300°C, evaluates the reactor's dynamics when subjected to thermal perturbations. The scenario for this transient is found in Figure 4a, and it represents a blockage in the feed-water system inside the HTR-Modul-200 [5].

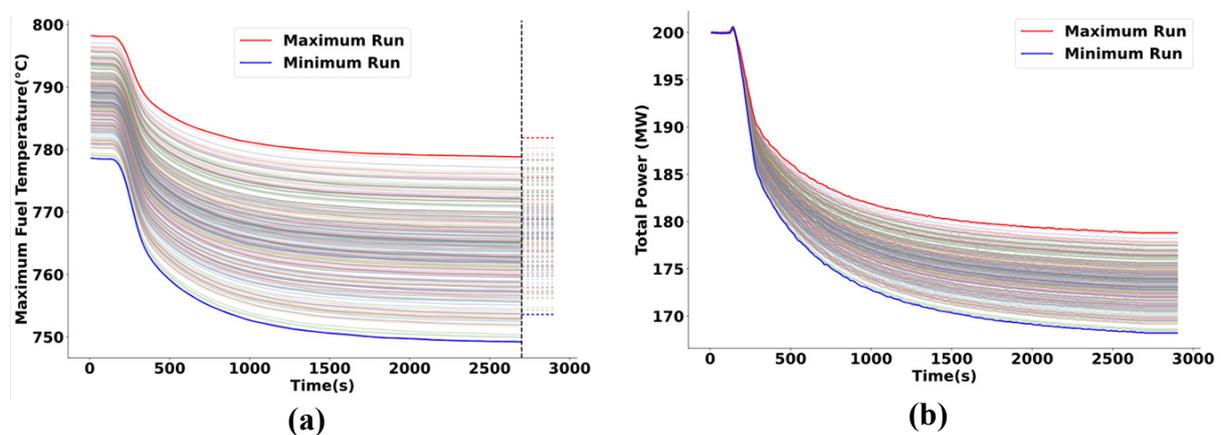

(a)      (b)

**Figure 8: Inlet Gas Temperature Surge Transient Response of (a) Maximum Fuel Temperature and (b) Reactor Power.**

Figure 8a shows the trajectories of maximum fuel temperature over time for the different 200 samples of the inlet gas temperature perturbations, while Figure 8b shows the nominal power response to the same perturbations. The reactor's nominal power initially does not change before the ramp since the variation in the inlet temperature (here due to the uncertainty) is accompanied with change in the maximum fuel temperature (and hence the outlet gas temperature) that will compensate each other in order to get constant nominal power of 200 MW. After 2 minutes, the inlet temperature is ramped from its initial value (which varies according to the applied uncertainty with a mean of 250ºC) to 300ºC which is constant for all samples. This increase in the gas inlet temperature will result in a reduction in the reactor's nominal power as observed in Figure 8b which also results in a reduction in the maximum fuel temperature. After around 45 minutes (2700 seconds), the reactor tends to stabilize, and the second steady state starts from the final transient conditions. The spread of the maximum fuel temperature at the end of the transient for all samples was computed as 22ºC while that of the power was roughly 5MW. This spread in the nominal power for the second steady state is due to having a constant inlet temperature of 300ºC among all the samples, while each sample had a different maximum fuel temperature (and hence outlet gas temperature) coming from the transient state. Figure 9 shows the mean and the variation due to the reactor's response to the inlet gas temperature.

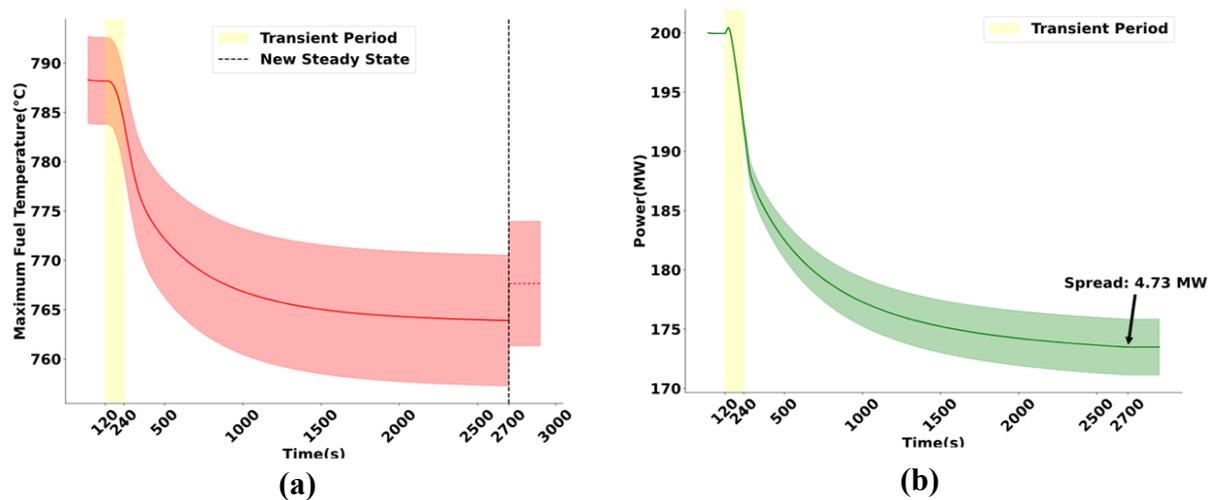

(a)            (b)

**Figure 9: Stabilization of (a) Maximum Fuel Temperature and (b) Total Nominal Power Post-Inlet Temperature Surge.**

The second transient simulated is the one shown in Figure 4b. The loss of forced cooling under pressure is attributed to the loss of the gas flow rate inside the media. The mass flow rate is reduced from its initial value to 0% over 30 second. Figure 10 captures the transient response in fuel temperature and reactor power due to this loss of forced cooling event. The large temperature gradient within the core coupled with the pressurized helium coolant can lead to strong natural circulation flows. These internal convective currents augment heat transfer from the fuel elements and improve core cooling, resulting in maximum fuel temperature not exceeding a peak value of 1105ºC among all the 200 samples simulated. This is well below the design limit of the fuel used inside HTR-Modul-200 which is around 1600ºC [5, 10].

The fuel temperature increases at the beginning of this accident due to the loss of the forced cooling, reaching a peak value but then decreases due to the reduction of the reactor power and the help of the natural convection established. The reactor will follow the decay power profile, as seen in the semi-log plot in Figure 10b, and enough decay heat will raise the fuel temperature a second time but albeit lower in value when compared to the initial loss of forced cooling event. This heat is dissipated effectively, and the fuel temperature continues to decrease. All the 200 samples have shown this behavior, and the relatively narrow spread in reactor power across the samples illustrates the system's uniform power reduction and the prompt reduction in reactor's power.

The histogram in Figure 11 centres around a mean peak temperature of ~1092ºC, with a standard deviation of around 5ºC, emphasizing a consistent peak temperature response in the different samples for the perturbation of the inlet gat temperature. The results show that the reactor can still be safely operated within the uncertainty limit on the inlet gas temperature. Such insights that one can take from Figures 10 and 11 provide a comprehensive view of the reactor's thermal behaviour under one of the important transient scenarios for HTGR, demonstrating the capacity and the importance of the UQ framework developed.

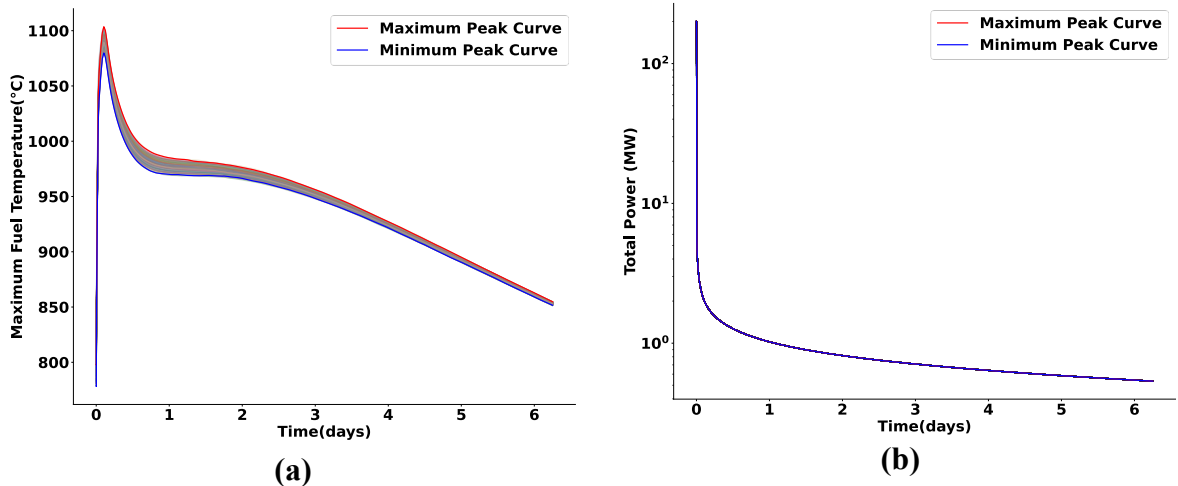

**Figure 10: Transient PLOFC Impact on (a) Fuel Temperature and (b) Reactor Power.**

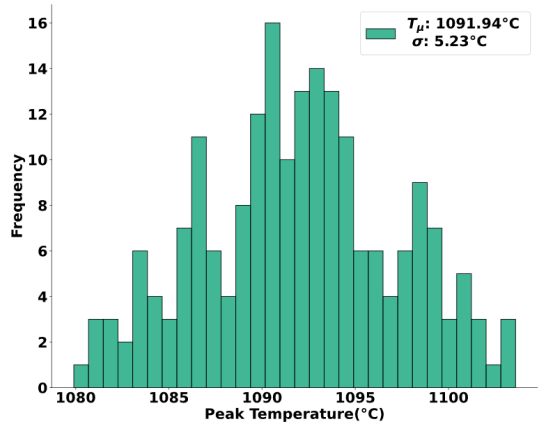

**Figure 11: Fuel Temperature Peak Distribution.**

### 3.3.3 *Neutronics case*

For the neutronics case study, we selected one example of each of manufacturing and material properties uncertainties to propagate through the model. Specifically, we chose U-235 fuel enrichment, which represents a manufacturing uncertainty, and graphite density in the reflector and matrix materials, which represents a material properties uncertainty.

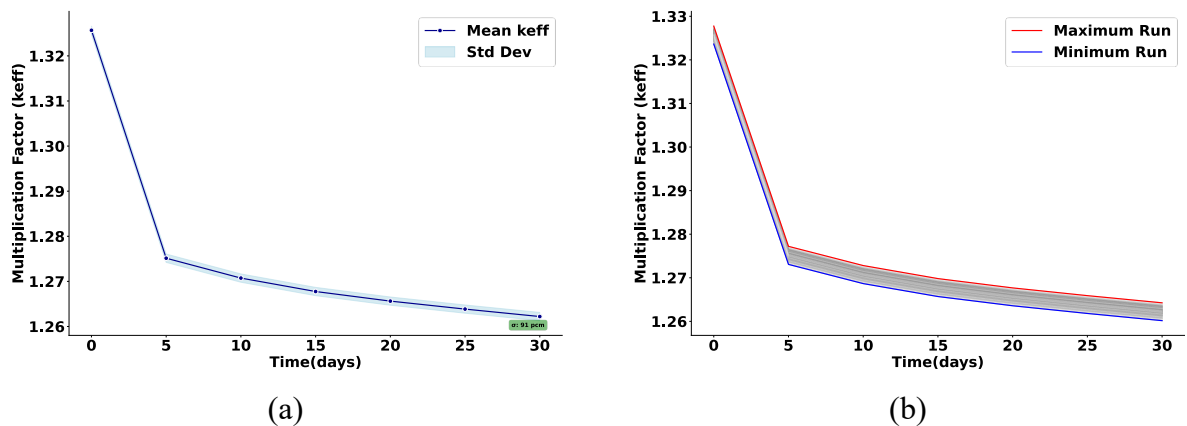

**Figure 12: Neutron Multiplication with Fuel Enrichment Variance. (a) Mean Multiplication Factor (b) the 200 Multiplication Factor Samples Runs.**

Figures 12a-13a illustrate how the multiplication factor varies based on the reported uncertainties from Table 2. The data in the figures depict a burnup duration of 30 EFPD for the active core of the HTR-Modul-200 reactor. The mean $k_{eff}$ decreases over time, characteristic of fuel burnup, and the standard deviation at the end of the 30 EFPD is 91 pcm and 15 pcm, respectively. Figures 12b-13b show the maximum and minimum $k_{eff}$ curves for the 200 samples simulated.

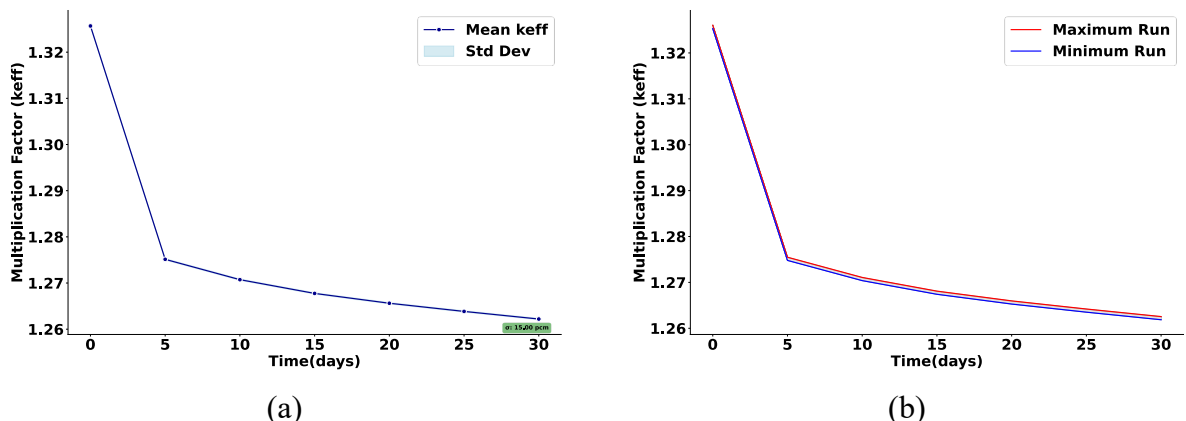

(a)                      (b)

**Figure 13: Neutron Multiplication with Graphite Density Variance. (a) Mean Multiplication Factor (b) the 200 Multiplication Factor Samples Runs.**

## 4. CONCLUSION AND FUTURE WORK

This study has demonstrated the integration of the High Temperature Reactor Code Package HCP with the DAKOTA toolkit, establishing a robust framework for uncertainty quantification in the modelling and simulation of HTGRs. Through application of this framework to the HTR-Modul-200 reactor, we have propagated manufacturing, material, and initial/boundary condition uncertainties in steady state and transient multiphysics cases. The insights provided in this paper pave the way for targeted efforts to reduce predictive uncertainty and enhance the robustness of HTGR systems.

For future work, there are several promising avenues for extending the capabilities of this integrated UQ framework. HCP's energy library generator, currently linked to NJOY [6], facilitates direct processing of isotope-wise data from ENDF nuclear data libraries. A preprocessing step involving perturbation of ENDF files using tools like SANDY [11] is envisioned to enhance the fidelity of nuclear data uncertainty sampling and propagate these uncertainties in HCP and HTGR simulation. Furthermore, HCP relies on porous media equations and HTGR experimental correlations to define several fluid dynamics and thermal hydraulics quantities used internally in the HCP source code (e.g. Nusselt number correlations, flow resistance estimation, thermal conductivities). These correlational models could be targeted for uncertainty analysis to propagate relevant uncertainties into the framework, benefiting HTGR modelling and simulation.

Beyond the current multiphysics analysis that couples neutronics and fluid dynamics and thermal hydraulics, HCP includes other specialized modules that could be incorporated into the uncertainty analysis framework for a more comprehensive treatment of HTGR modeling. For example, the STACY module simulates fission product release, while the STAR module analyzes dust particles resuspension and inclusion into source term analysis, and SHUFLE handles fuel shuffling capabilities. By propagating uncertainties through these different HCP components focused on specific physical phenomena, the overall credibility and robustness of HTGR simulation can be strengthened. HCP represents a vast modeling code with advanced capabilities spanning multiple aspects of HTGR behavior. Applying uncertainty quantification across HCP's extensive features sets for integrated multiphysics analysis would further leverage the power of this software and framework to enable robust predictions and design optimization for HTGRs. By advancing these technical components, the HCP and DAKOTA integration will be better positioned to support the nuclear industry's uncertainty management activities and will contribute to the enhanced reliability of HTGR designs and support their licensing and development.